# Go Heuristics for Coverage of Dense Wireless Sensor Networks

**M. A. El-Dosuky [1], M. Z. Rashad [1], T. T. Hamza [1], and A.H. EL-Bassiouny [2]**

[1] Department of Computer Sciences, Faculty of Computers and Information sciences, Mansoura University, Egypt

[2] Department of Mathematics, Faculty of Sciences, Mansoura University, Egypt

**Abstract**

This paper collects heuristics of Go Game and employs them to achieve coverage of dense wireless sensor networks. In this paper, we propose an algorithm based on Go heuristics and validate it. Investigations show that it is very promising and could be seen as a good optimization.

## 1. Introduction

Integration and scalable coordination of many sensor nodes in a wireless sensor network (WSN) [11] is among this century challenges ([1], [2]). To provide energy-efficient communication protocol and coordination algorithm for topology maintenance, this requires employing global optimization of **communication** and **sensing** ([5], [6]).

In designing a sensor network, there are certain QoS parameters that need to be considered, such as density, accuracy, latency and lifetime [3]. We consider density as the most important parameter to fulfill network coverage with energy saving ([4] [5]).

We review density in section 2, and then introduce heuristics of Go Game in section 3.

## 2. Density and Coverage

Density λ is defined as the average number of neighbors per node:

$$\lambda = \frac{N\pi R^2}{A} \qquad (1)$$

where $R$ is node range, $A$ is area of sensor field, and $N$ is total number of deployed nodes.

Edge affects are ignored, assuming the network is connected with very low probability of isolated node [9]:

$$P(isolated\ node) = (1 - e^{-\lambda})^N \qquad (2)$$

Connectivity of wireless sensor networks is analyzed to show if the probability of isolation is enough to represent connectivity [7].

Coverage may be defined as quality of service [13]. It relates to the capacity of the wireless networks [15] and the exposure [14].

## 3. Go Heuristics

Go is an Asian game especially in China, Japan and Korea, now in the rest of the world. The game is played on a board with a grid of 19x19 intersections. Black and white players can put a single stone on any unoccupied intersection. A player can pass at any turn. A ***liberty*** is a gap or an empty intersection. Liberties are shared amongst connected stones. A stone is captured when the last of its liberties is removed [16]. Figure 1 shows an example of the white capturing by playing at any triangle. .

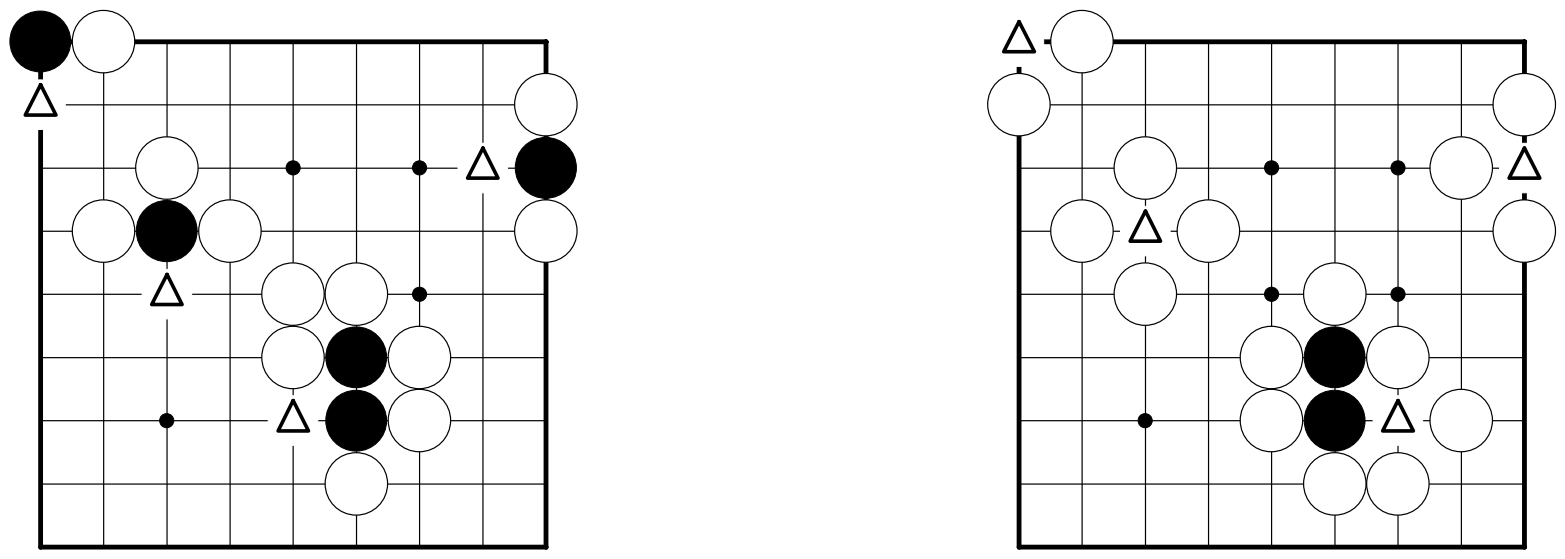

a) before b)after

**Figure 1 white Captures black** [*18*]

Artificial intelligence algorithms are applied to simulate human Go playing techniques [17]. A human player applies many strategies and tactics that can be easily programmed such as [*18*]: keep your stones connected to each other, avoid making lots of groups, Look for weak groups, i.e, connected to relatively few empty intersections, and try to surround empty intersections. Based on these four heuristics, which we can call Hs, we propose the algorithm, shown in figure 2.

Note that the last variable, threshold, is tolerable error. It can be fixed to 0.1, or determined after reviewing critical density thresholds in distributed wireless networks [8].

**function GO_HEURISTICS** (*board,* R, A)

**input:**

*board*, a set of intersections ($x_i$, $y_i$)

R , node range

A is area of the sensor field,

**Variables:**

```
        λ, density
        p, P(isolated node)
        N, total number of deployed nodes, initially =1
        closed, set containing all occupied points, initially empty
        current, current intersection point
        Hs, set of Heuristics
        threshold, tolerable error, fixed to 0.1
    begin
        λ = NπR^2 / A ,          p = (1 − e^{−λ})^N
        while p < threshold do
            current = RANDOM_SELECT(board, Hs)
            if current ∉ closed then
                closed = closed ∪ { current }
                N = N+1
                λ = NπR^2 / A ,          p = (1 − e^{−λ})^N
            end if;
        end while;
    end
```

**Figure 2. Pseudo code for GO_HEURISTICS.**

## 4. Evaluation

A common case study for benchmark analysis is of ultrasonic sensors with the following parameters [10]:

Area is 100x100m. The sensing range of the nodes is 7m. Using this we get the effective node density from sensing perspective as

$$\lambda = \frac{N\pi R^2}{A} \qquad (3)$$

The probability that a unit area is in the range of *n* nodes is given by

$$P(n) = P_R^n (1-P_R)^{N-1-n} \binom{N-1}{n} \quad (4)$$

$$\text{where } P_R = \frac{\pi R^2}{L^2} \quad (5)$$

This binomial distribution converges towards the Poisson distribution, for large values of $N$ [12] :

$$P(n) = \frac{e^{-\lambda_s} \lambda^n}{n!} \quad (6)$$

Function is increasing with number of nodes and would be normalized to 1. As shown in figure 3.

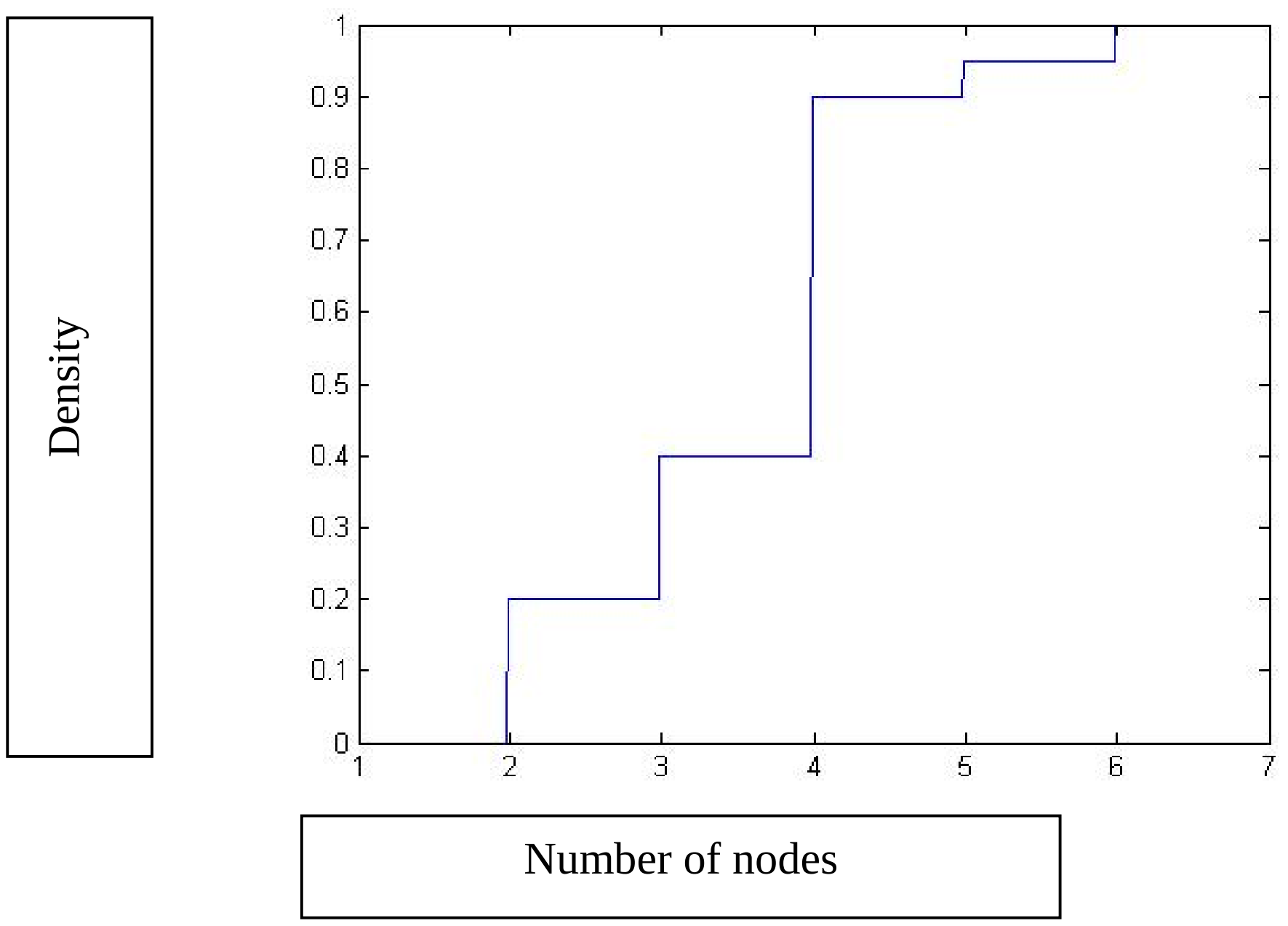

Figure 3: Shaping function